\documentclass[onecolumn]{aastex631}
\usepackage{graphicx} 
\usepackage{amsmath}
\usepackage[]{natbib}
\usepackage{color}

\graphicspath{{./}}

\begin{document}

\title{In-depth analysis of bar formation mechanisms of disk galaxies in halos of different concentrations}
\author{T. Worrakitpoonpon}
\affiliation{Institute of Science, Suranaree University of Technology, Nakhon Ratchasima 30000, Thailand}

\correspondingauthor{T. Worrakitpoonpon}
\email{worraki@gmail.com}
\date{}

\begin{abstract}
We use $N$-body simulations to investigate the distinct bar formation processes in disks residing in halos of various concentrations. In a highly concentrated halo, the bar development is limited by the dominant multi-arm modes as a result of the swing amplification in the early stage. Only after the multi-arm modes decay, the bar growth proceeds mechanically owing to the particle trapping in continuation of that bar seed. In this scheme, the corotation resonance of the bar modes does not come into play at all, justified by a low amount of disk-halo angular momentum transfer and a modestly decreasing bar pattern speed. On the other hand, although reducing the halo concentration suggests the reduction of the preferred swing-amplified modes to be bi-symmetric, the bar formation in a lowly concentrated halo does not involve the swing amplification at all. Rather, the fast-growing linearly unstable bar modes of a single uniform frequency is solely the governing factor, attributed to a mild shearing. The bar modes trigger the corotation resonance since the beginning and such resonance is maintained until the end, which leads to a high amount of angular momentum transfer and a fast slowdown. For the intermediate halo concentration, the kinematical analyses of multiple non-axisymmetric modes suggests that the linear modes, the swing amplification, and the particle trapping are all present in the evolution chronology.
To specify bars formed in the different halo concentrations, full analyses of the isophotal shape, the radial Fourier amplitude, and the resonance diagram can be of use.

\end{abstract}

\section{Introduction}
\label{sec:intro}

Barred structure is a common feature found in more than $60\%$ of disk galaxies nowadays \citep{eskridge_et_al_2000,lee+ann+park_2019} and its existence was spotted at as far as the redshift $2$, although the fraction of the bar-hosting disk decreased to $\sim 10\%$ at such distance \citep{kuhn_et_al_2024,espejo_salcedo_et_al_2025}. That redshift limit has been uplifted by the high-resolution JWST survey which was able to pinpoint bars in galaxies of redshift $4$ \citep{smail_et_al_2023,amvrosiadis_et_al_2025}.
Those observations led to subsequent analyses which were not limited to the bar morphology \citep{carillo_et_al_2020,tahmasebzadeh_et_al_2021} and kinematics \citep{jimemez-arranz_et_al_2024,geron_et_al_2024}, but the physical processes and properties beneath such as the star formation \citep{george_et_al_2020,kim_et_al_2024sfr}, the gas concentration \citep{gonzalez-alfonso_et_al_2021,yu_et_al_2022}, or the stellar age \citep{fraser-mckelvie_et_al_2019,neumann_et_al_2024}, remained active topics under investigation.

High prevalence of the bars urged astrophysicists to venture into understanding its origin, and the bar formation formalisms could be bifurcated into two paradigms. The first framework hypothesized the spontaneous bar formation in a bar-unstable disk, stemming from the global unstable non-axisymmetric modes triggering the bar growth \citep{hunter_1963}, known as the internal or secular scheme. That conjecture has been pioneeringly verified in simulations of an isolated disk of particles \citep{hohl_1971} and a particulate disk-halo system \citep{sellwood_1980}. These led to numerous subsequent numerical investigations of the evolution of disk systems of various properties, in parallel with the refinement of the numerical techniques for simulations (see, e.g., \citealt{valenzuela_et_al_2003,ceverino+klypin_2007,foyle_et_al_2008,zou_et_al_2014,pettitt+wadsley_2018,sellwood_et_al_2019,fujii_et_al_2019,collier+madigan_2021,valencia-enriquez_et_al_2023,ghosh_et_al_2023,joshi+widrow_2024}).
The other scenario viewed the bar formation differently as being caused by the galaxy encounter, namely, the external scheme. In this respect, the disks were perturbed by the tidal forces from the fly-by to establish a bar, as originally verified in a simple suite of simulations \citep{toomre+toomre_1972}. Although this hypothesis has been formulated long before, the interest on it has been upraised only recently owing to the accessibility to the large-scale realistic $\Lambda$CDM cosmological simulations, which allowed us to investigate closely the interactions between galaxies and environmental effects constituting a bar \citep{zhao_et_al_2020,reddish_et_al_2022,cavanagh_et_al_2022,bi_et_al_2022,fragkoudi_et_al_2025,ansar_et_al_2025,chim_ramirez_et_al_2025,lu_et_al_2025}. The two frameworks were long considered as the two distant paradigms and were usually investigated separately until recent studies have performed the unified investigations for the census \citep{zana_et_al_2018,lokas_2021bar}, the underlying mechanism \citep{zheng+shen_2025}, and the timescale \citep{zheng_et_al_2025}, and conclusions were drawn between them. 

Focusing on the secular scope, the bar instability and the resulting bar depended on many factors. The central mass concentration affected significantly the bar strength \citep{athanassoula_2003}, the bar length \citep{michel-dansac+wozniak_2006,petersen_et_al_2024}, the formation timescale \citep{curir_et_al_2007,polyachenko_et_al_2016,jang+kim_2023,bland_hawthorn_et_al_2023,worrakitpoonpon_2025}, the bar pattern speed \citep{kataria+das_2019}, and the stability against bar modes \citep{shen+sellwood_2004,saha+elmegreen_2018,fujii_et_al_2018,kataria+das_2018}. On the other hand, the studies of the effect from the disk velocity dispersion, conventionally parameterized by the Toomre's $Q$ parameter \citep{toomre_1964}, covered a smaller scope as it was mainly related with the timescale and the stability \citep{jang+kim_2023,worrakitpoonpon_2023,cas_paper}.

Despite that the bar instability has been extensively investigated for decades, the understanding on detailed processes of the bar formation in the different ranges of parameters remains far from completion. When the question of the bar instability arose, such process was regularly connected with a number of sub-mechanisms. However, a unified picture---where and when each process takes the role---is still incomplete. In this study, we enlarge the scope in the way that the physical processes underlying the bar formation and evolution for different initial states are carefully analyzed in the attempt to find the favorable condition for each process to take the role.
Before we step into the main context of this article, we recall the important theoretical milestones associated with the formation of the bar, or more generally the non-axisymmetric structures. The first one is the linear theory which hypothesized that all non-axisymmetric $m$-armed modes were embedded in the disk of particles since the start with the density described by  
\begin{equation}
\delta\Sigma \sim e^{i(m\theta+\omega' t)}. \label{eq:d_sigma}
\end{equation}
This represented the $m$-armed perturbations of the uniform frequency $\omega '$ which became linearly unstable if $\omega'$ was imaginary \citep{hunter_1963,kalnajs_1971}, triggering the exponential growth supported by the resonance. It was suggested that the linear bar (or $m=2$) modes were always the dominant over the other ones. Otherwise, the formation of the non-axisymmetric $m$-armed modes could be regulated by the swing amplification \citep{julian+toomre_1966}, conjecturing that the spiral arms or the bar were the consequence of the disk response, or the wake, to the disturbance from the combined shearing/epicyclic motion. The preferred number of the arms $m$ can be estimated by the circumferential wavelength of the wake  \citep{toomre_1981,sellwood+carlberg_2014} that yields
\begin{equation}
m\sim \frac{R_{CR}\kappa^{2}}{4\pi G\Sigma},
    \label{eq:arm_num}
\end{equation}
based on the assumption that the swing amplification was most efficient around the corotation radius $R_{CR}$. From Eq. (\ref{eq:arm_num}), increasing $\kappa$ by increasing the central mass concentration led to a higher $m$. 
Lastly, we recall the theory of the orbital or particle trapping. It has been pioneered by \citet{lynden-bell_1979trap} who considered an adiabatically invariant action variable centered in the rotating frame of frequency
\begin{equation}
\Omega_{i}=\Omega-\frac{\kappa}{2},
    \label{eq:omega_i}
\end{equation}
where $\Omega$ and $\kappa$ are the orbital and epicyclic frequencies, respectively. Based on that assumption, the orbit whose orbital frequency slightly mismatched with the perturbation angular frequency, namely, the bar pattern speed, tended to re-align with that bar. That conclusion was in line with \citet{contopoulos+papayannopoulos_1980} who discovered that particles constituting a bar originated from a stable family that co-rotated with the bar along the bar axis, verified by orbital integrations. In the dynamical perspective, both works suggested that particle's angular frequencies were able to be mechanically trapped by and co-rotate with the bar potential, strengthening the bar in process.
The two first frameworks were frequently considered in explaining the formations of the bar and the spiral arms in various simulations, whereas in this study we will not only analyze the interplays between those two mechanisms but the third mechanism, with a proper indicator, will be addressed as well. Some studies specified the growth of the non-axisymmetric modes triggered by a strong perturber, such as a heavy point mass \citep{kumamoto+noguchi_2016,sellwood+carlberg_2021}, as the non-linear modes. These ones are out of our scope as we focus on the disk without preset initial perturbations; the initial finite-$N$ noises fully originate from the randomly placed particles.

The article is organized as follows. First, Sec. \ref{sec:sim_param} provides details of the employed disk model and the simulation setting. Next, Sec. \ref{sec:nume} reports the results of the morphological and kinematical analyses of the bar formation in different cases and the connections with observations. It is then followed by Sec. \ref{sec:shear_stab} that exemplifies the dynamical analysis of the disk interior responsible for the different bar formation schemes. Finally, the studies are concluded in Sec. \ref{sec:conclu}.

\section{Initial disk model and simulation details}
\label{sec:sim_param}

The system for investigation is of the disk of particles in the live halo. The disk follows the three-dimensional exponential density profile given by
\begin{equation}
\rho_{d}(R,z)=\frac{M_{d}}{4\pi R_{0}^{2}z_{0}}e^{-R/R_{0}}\text{sech}^{2}{\bigg(\frac{z}{z_{0}}\bigg)}
    \label{density_disk}
\end{equation}
where $M_{d}$ is the disk mass, $R_{0}$ is the disk scale radius, and $z_0$ is the disk scale thickness, fixed to $9.3\times 10^{9} \ M_{\odot}$, $5 \ \text{kpc}$, and $0.2 \ \text{kpc}$, respectively. The disk is radially and vertically truncated at $5R_{0}$ and $5z_{0}$. The spherically symmetric halo has the Hernquist density profile \citep{hernquist_1990}
\begin{equation}
\rho_{h}(r)=-\frac{M_{h}r_{h}}{2\pi r(r+r_{h})^{3}},
    \label{den_hern}
\end{equation}
where $M_{h}$ and $r_{h}$ are the halo mass and the halo scale radius. The halo mass is fixed to $25M_{d}$ whereas we vary $r_{h}$ from $35-75 \ \text{kpc}$ for the different levels of the central mass concentrations. The halo of particles is truncated at $2r_{h}$. 

The initial radial $Q$ profile corresponds to the ratio of the radial velocity dispersion $\sigma_{R}$ to the critical value according to the Toomre's criterion \citep{toomre_1964,nipoti_et_al_2024}, i.e.,
\begin{equation}
    Q=\frac{\sigma_{R}\kappa}{3.36G\Sigma}
    \label{q_def}
\end{equation}
where $G$ is the gravitational constant; $\kappa$ is the epicyclic frequency calculated from the total disk-halo potential; and $\Sigma$ is the disk surface density which is proportional to $e^{-R/R_{0}}$. The velocity moments are chosen from the prescriptions of \citet{hernquist_1993}. The radial velocity dispersion relates with $\Sigma$ as
\begin{equation}
\sigma_{R}^{2}\propto \Sigma, \label{sigmar_prop}
\end{equation}
and the minimum $Q$ (or $Q_{min}$), numerically calculated from Eq. (\ref{q_def}), represents the initial kinematical state. This choice of $\sigma_{R}$ yields the characteristic U-shape radial $Q$ profile with $Q_{min}$ located close to $2R_{0}$ (see \citealt{worrakitpoonpon_2025} for details). The tangential velocity dispersion profile $\sigma_{\theta}$ is given by
\begin{equation}
\sigma_{\theta}^{2}= \frac{\kappa^{2}}{4\Omega^{2}}\sigma_{R}^{2}, \label{sigmatheta}
\end{equation}
where $\Omega$ is the angular frequency of the circular orbit computed from the total potential, and the vertical velocity dispersion $\sigma_{z}$ is obtained from 
\begin{equation}
\sigma_{z}^{2}= \pi G\Sigma z_{0}. \label{sigmaz}
\end{equation}
The mean tangential velocity $\bar{v}_{\theta}$, i.e., the rotation curve, is deduced from the axisymmetric Jeans equation
\begin{equation}
\bar{v}_{\theta}^{2}=R^{2}\Omega^{2}+\frac{R}{\Sigma}\frac{d(\sigma_{R}^{2}\Sigma)}{dR}+\sigma_{R}^{2}-\sigma_{\theta}^{2}.
\label{vthetamean}
\end{equation}
The mean radial and vertical velocities, namely, $\bar{v}_{R}$ and $\bar{v}_{z}$, are initially zero. For the halo, the first moments of the velocities are zero while the velocity dispersion $\sigma_{h}$, supposedly isotropic, is numerically determined from the spherically symmetric Jeans equation, arranged to be in the form
\begin{equation}
    \sigma_{h}^{2}=\frac{1}{\rho_{h}(r)}\int_{r}^{\infty}\frac{GM_{tot}(r)\rho_{h}(r)}{r^{2}}dr,
\label{sigma_halo}    
\end{equation}
where $M_{tot}(r)$ is the total mass inside $r$. Disk and halo random velocity components are drawn from the cut-off Gaussian distribution. 

The disk and the halo are constituted from $2\times 10^{6}$ and $3\times 10^{6}$ particles, respectively. Disk-halo dynamics is simulated by GADGET-2, a publicly available code for astrophysical/cosmological simulations \citep{springel_2005}. For the force calculation, we fix the softening length to $5$ pc and the opening angle to $0.7$, applied to all particles. The integration time step is controlled to be below $0.1 \ \text{Myr}$ so that the deviations of the total energy and the total angular momentum at the end of simulation are not greater than $0.2 \%$ of the initial values. We investigate the disks with $Q_{min}=1.1$ for five different $r_{h}=35, 40, 50, 60$ and $75 \ \text{kpc}$. We focus on the $Q_{min}$ close to unity to ensure that the bar formation is accomplished within the simulated timescale as a higher $Q_{min}$ result in a slower bar formation or stability. Some studies further parameterized the degree of the central mass concentration by the ratio of the mass enclosed in a specific radius to the total mass for specific analytical purposes (see, e.g., \citealt{jang+kim_2023, worrakitpoonpon_2025}). In this study, the central mass concentration is considered only in the relative sense such that the variation from low to high $r_{h}$ signifies the high to low central mass concentration for interpreting the results. 
Each case is nominated by the value of $r_{h}$ after the letter 'R'. For instance, the case name R50 stands for the case with $Q_{min}=1.1$ and $r_{h}=50 \ \text{kpc}$.

\section{Analysis of disk evolution in simulations}
\label{sec:nume}

\subsection{Development of non-axisymmetric features for different $r_{h}$}
\label{ssec:evo_morpho}

We first of all inspect the morphological evolution of the disks in the course of the bar formation and the evolution after for various $r_{h}$. Non-axisymmetric $m$-mode features of a disk of particles can be numerically evaluated by the $m$-mode Fourier amplitude as a function of radius $\tilde{A}_{m}(R)$ defined as
\begin{equation}
    \tilde{A}_{m}(R)=\frac{\sqrt{a^{2}_{m}+b^{2}_{m}}}{A_{0}}
    \label{tilde_a2}
\end{equation}
where $a_{m}$ and $b_{m}$ are the $m$-mode Fourier coefficients as a function of radius and $A_{0}$ is the corresponding $m=0$ amplitude. The $m$-mode strength $A_{m}$ is designated by the maximum $\tilde{A}_{m}$ within $R_{max}=10$ kpc. We employ $A_{2}$ as the bar strength and the disk is considered fully barred when $A_{2}$ exceeds and remains above $0.2$. The $m=2$ phase as a function of radius $\phi_{2}(R)$ can be calculated by
\begin{equation}
    \phi_{2}(R)=\text{atan2}(b_{2},a_{2}),
    \label{phi2_r}
\end{equation}
which is useful for inspecting the particle dynamics in response to the bar modes at different radii. Otherwise, the bar phase $\phi_{2,tot}$ is computed in the same way as in Eq. (\ref{phi2_r}) but the Fourier coefficients are determined from all particles inside $R_{max}$. The angular motion of the bar is investigated using the latter phase. 

\begin{figure*}
    \centering
    \includegraphics[width=18.0cm]{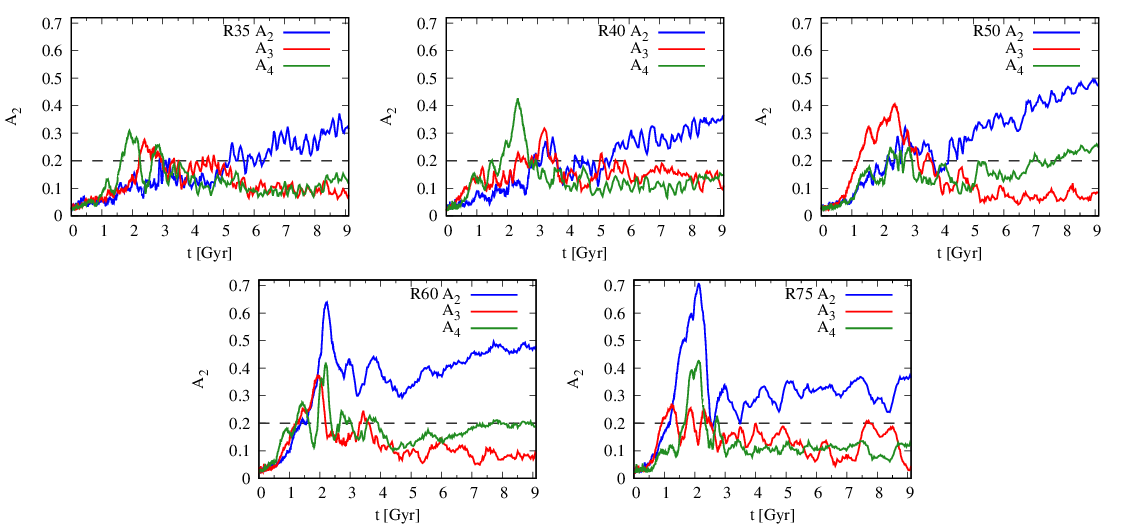}
    \caption{Time evolutions of $A_{2}, A_{3}$ and $A_{4}$ for different indicated cases. The horizontal dashed line designates the value of $0.2$, serving as the threshold of $A_{2}$ for the barred state.}
    \label{fig:a2}
\end{figure*}

\begin{figure*}
    \centering
    \includegraphics[width=18.0cm]{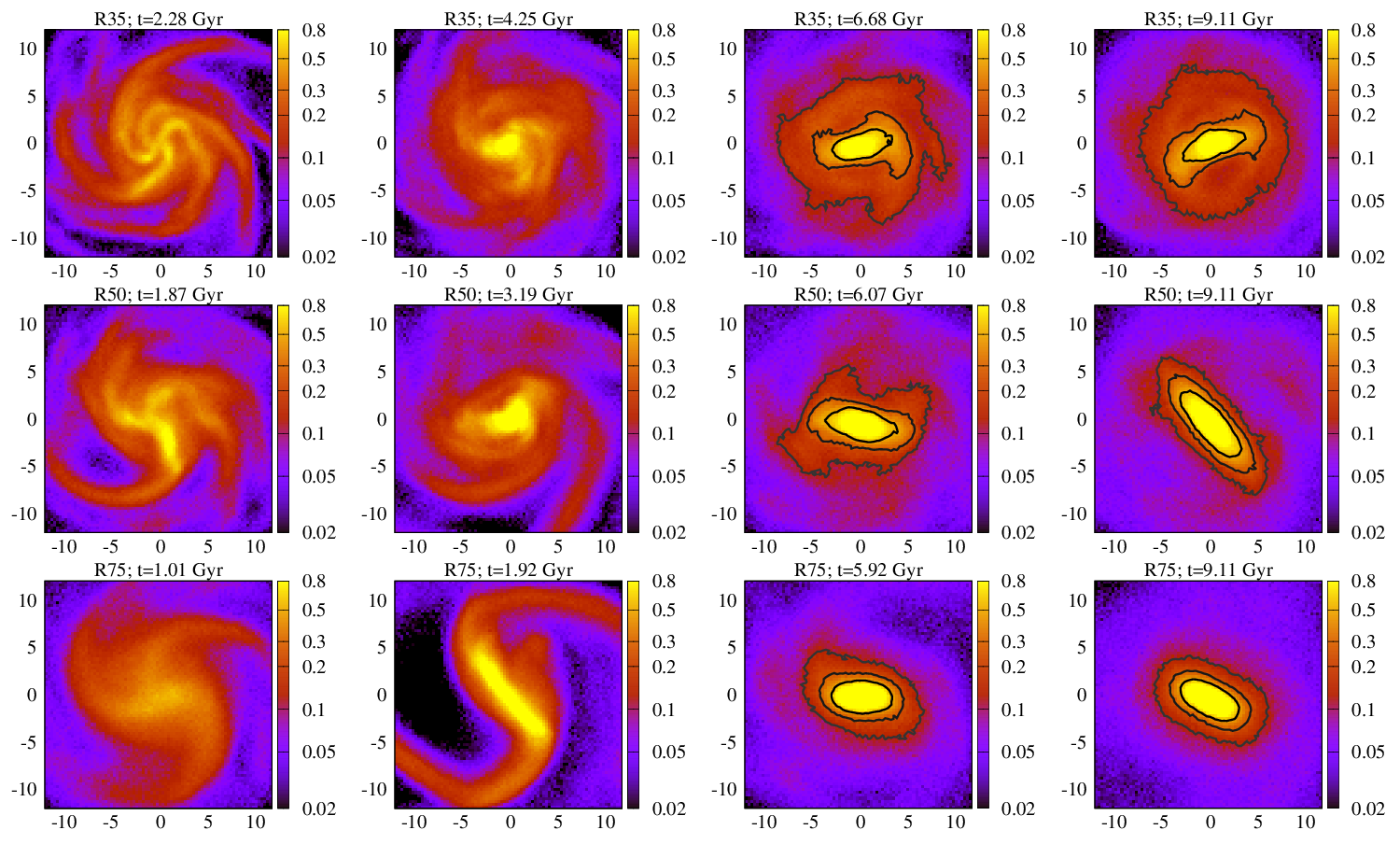}
    \caption{Disk surface densities in color map for R35 (top row), R50 (middle row), and R75 (bottom row) at different times, in units of $9.3\times 10^{7} \ M_{\odot}/$kpc$^{2}$. The isodensity contours in the last two snapshots represent, from the outside to the inside, the levels of $0.2\Sigma_{0}, 0.4\Sigma_{0}$ and $0.8\Sigma_{0}$, where $\Sigma_{0}\equiv M_{d}/2\pi R_{0}^{2}$ corresponds to the central surface density of the initial disk.}
    \label{fig:snap}
\end{figure*}

Shown in Fig. \ref{fig:a2} is the evolution of $A_{2}, A_{3}$ and $A_{4}$ for all cases investigated. In accompanying the evolutions of the three modes, the morphologies of R35, R50, and R75, representing the disks in high, intermediate, and low halo concentrations at different times, are depicted in Fig. \ref{fig:snap} with the isodensity contours of different levels in the two last snapshots. The two plots exhibit different schemes of the bar formation for different $r_{h}$. For the two lowest $r_{h}$, equivalent to the two most concentrated halos, the multi-arm modes, designated by $A_{3}$ and $A_{4}$, rise more rapidly and dominate the $m=2$ modes in the first $3 \ \text{Gyr}$. The formation of the spiral arms is attributed to the swing amplification \citep{julian+toomre_1966}. This topic, along with the other mechanisms forming and evolving the bar, will be discussed more in Sec. \ref{ssec:kine_bar}.
The multi-arm modes start to decay at $3 \ \text{Gyr}$, at the time the bar modes start growing uninterruptedly. These are in line with the configuration of R35 in Fig. \ref{fig:snap} as the disk has the multi-arm pattern at $2.28 \ \text{Gyr}$ and the spiral arms start to dissolve at $4.25 \ \text{Gyr}$. Afterward, the bar becomes increasingly prominent, as evaluated by the $0.4\Sigma_{0}$ isodensity lines in the last two snapshots. 
It turns out that in the highly concentrated halo regime, the bar growth in the midst of the swing-amplified spiral arms is limited by that cause. Nevertheless, the significant bi-symmetric features, as indicated by $A_{2}$ that almost reaches $0.2$ albeit the dominance of the $m>2$ modes, is already developed. After the drops of $A_{3}$ and $A_{4}$, this bar seed promptly grows and surpasses $0.3$ at the end. The fact that the bar modes dominate the long-term evolution can be explained that the spiral modes are the short-lived modes, fading away within multiple epicyclic periods as estimated by \citet{julian+toomre_1966,michikoshi+kokubo_2016}. 
On the contrary, the global bar modes are growing exponentially \citep{sellwood+athanassoula_1986,dubinski_et_al_2009}, thus they overtake the spiral modes at the long race.

The formation scenario for the two highest $r_{h}$, namely, the diluted-halo limit, is different as the three modes are growing fast since the start but the $m=2$ modes are fastest-growing and tower above the others within $2 \ \text{Gyr}$. After $A_{2}$ reaches the peak, it relaxes and saturates around $0.3-0.4$, as with the other modes that relax and saturate well below $0.2$. The configuration of R75 at $1.01 \ \text{Gyr}$ is in agreement with the early evolution of $A_{m}$ as it demonstrates several weak non-axisymmetric modes before the two-armed modes take the dominance at $1.92 \ \text{Gyr}$. This produces the two-armed structure that is highly contrast to the background. Afterward, the bar is shortened in line with the relaxation of $A_{2}$ from the peak and it remains in size and shape until the end. Compared to the cases before, the bar strength is more saturated after $4 \ \text{Gyr}$. If evaluated by the $0.2\Sigma_{0}$ isodensity contour, the bar of R35 is embedded in a denser and more circularly dispersed background. 

The bar formation in the R50 case exhibits combined features such that the $m=3$ modes exert the dominance over the bar modes at the early time. Then, the bar development progresses continually, as for the low-$r_{h}$ regime. The bar at the final snapshot more resembles to that of the R75 as the $0.2\Sigma_{0}$ line is more oval in shape, indicating a more dominance of the $m=2$ modes than those in the R35 counterpart.
With a closer inspection on the evolutions of the three modes, the two fastest-growing modes change while we vary $r_{h}$. For R35 and R40, $A_{4}$ and then $A_{3}$ are the first and the second fastest-growing parameters, whereas they are altered to $A_{3}$ and $A_{2}$ for R50. The two cases with $r_{h}\geq 60 \ \text{kpc}$ are dominated by $A_{2}$ which reaches a highest peak. These are the crucial information before we conduct more the kinematical analysis in the next section.

In continuity with the inside-out variation of the shape of the isodensity contours spotted in the last snapshot of Fig. \ref{fig:snap}, we propose the ratio of the ellipticities of the two outermost contours that can potentially reflect the different formation schemes, namely, 
\begin{equation}
E = \frac{e_{0.2}}{e_{0.4}},
\label{eq:ratio_iso}
\end{equation}
where $e_{0.2}$ and $e_{0.4}$ are the ellipticities of the $0.2\Sigma_{0}$ and the $0.4\Sigma_{0}$ isodensity lines, respectively. As a consequence, we obtain $E=0.112,0.941,$ and $0.801$ for R35, R50, and R75, respectively. The ratio $E$ for R35 clearly stands out of the three cases, thus it is a useful identifier of the bar formed in a highly concentrated halo that resides in the circular dense background. 

\begin{figure*}
    \centering
    \includegraphics[width=18.0cm]{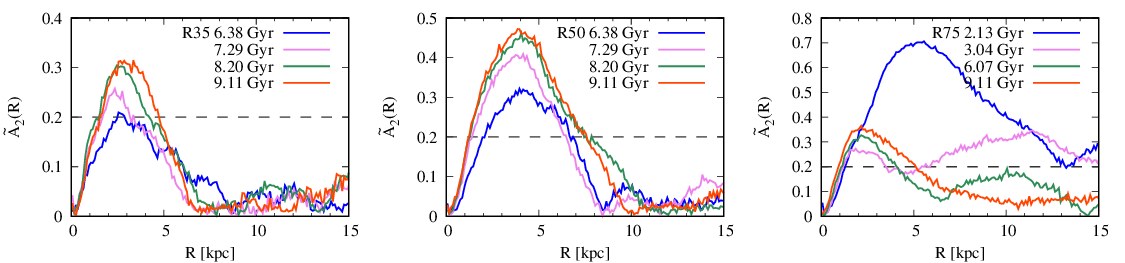}
    \caption{Radial profiles of the bar amplitude $\tilde{A}_{2}(R)$ for R35 (left panel), R50 (middle panel), and R75 (right panel) at the indicated times. The horizontal line indicates the value of $0.2$.}
    \label{fig:a2b2r}
\end{figure*}

The different mass distribution surrounding the bar can be alternatively investigated by the radial profile of the bar amplitude $\tilde{A}_{2}(R)$ for the three cases at the different times, plotted in Fig. \ref{fig:a2b2r}. For R35, a slow growth of the peak of $\tilde{A}_{2}(R)$ is spotted in the last $3$ Gyr, while the shape does not much alter. Beyond the sharp cut-off radius, $\tilde{A}_{2}(R)$ is close to the noise level until $15 \ \text{kpc}$, which indicates that the bi-symmetric modes beyond the bar are not prominent. The R75 counterpart exhibits substantial change of the shape in the course of the bar growth and the saturation afterward. At the end, $\tilde{A}_{2}(R)$ inside $5 \ \text{kpc}$ is similar to that for R35 but there is no clear cut-off. Rather, it saturates at the amplitude of $\sim 0.1$ until $15 \ \text{kpc}$. For R50, although the bar morphology at the final snapshot more resembles to that of the R75 case, the final $\tilde{A}_{2}(R)$ is more similar to the R35 counterpart such that $\tilde{A}_{2}(R)$ drops down to almost zero and remains at that level beyond.

The analyses of both the shape of the isodensity line and the form of $\tilde{A}_{2}(R)$ can effectively identify the bar originating from different central mass concentrations. Not only for the R35 and R75 bars of which the two profiles are definitely distinct, the bar of R50 can also be specified by the mixed features. The causes of the different morphological features spotted in both the isodensity lines and the $\tilde{A}_{2}(R)$ plots are attributed to the different bar formation mechanisms. These issues will be discussed more in Sec. \ref{ssec:kine_bar}.

\subsection{Dynamics and kinematics accompanying the bar formation and evolution}
\label{ssec:kine_bar}

\begin{figure*}
    \centering
    \includegraphics[width=18.0cm]{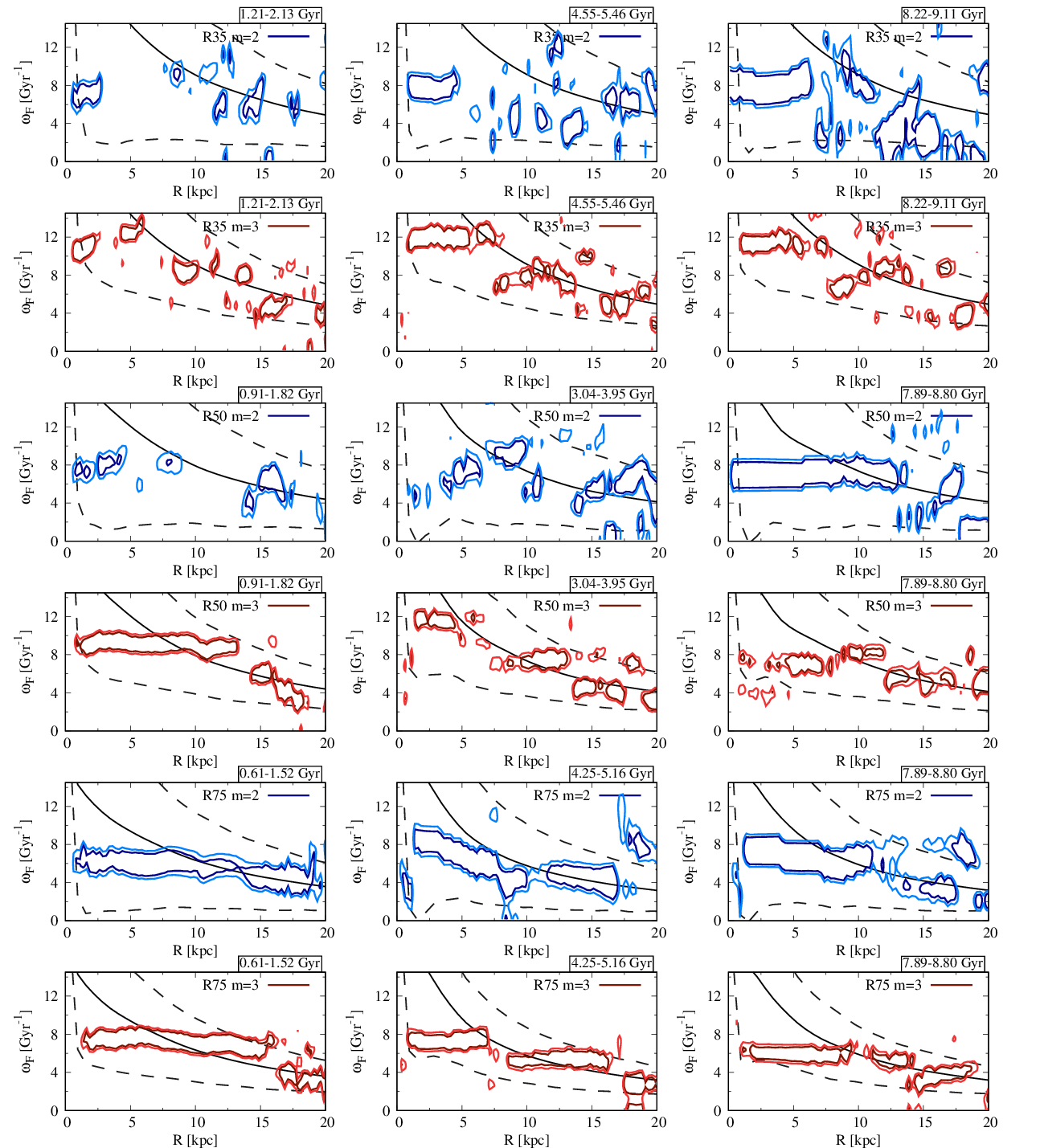}
    \caption{Fourier spectrograms of the $m=2$ and $3$ modes for R35, R50, and R75 (see labels), calculated in the time windows indicated on each panel. The Fourier amplitude is represented by the isolines of the levels of $0.7$ and $0.6$ times the maximum amplitude in the disk interior, from darker to lighter color for each mode. The solid and dashed lines correspond to the numerically computed disk angular frequency $\Omega$ and $\Omega\pm\kappa/m$, respectively, determined within the indicated time windows.}
    \label{fig:fourier_spec}
\end{figure*}

In continuity with the different bar formation scenarios in Sec. \ref{ssec:evo_morpho}, we further scrutinize the kinematical aspects of the different procedures. We consider first the kinematical evolutions of the $m=2$ and $3$ modes, demonstrated by the Fourier spectrograms $\omega_{F}$ at different epochs in Fig. \ref{fig:fourier_spec} with the $\Omega$ and $\Omega\pm\kappa/m$ profiles therein. The $m=3$ modes are the representative of the multi-arm modes since the overall evolution of the $m=4$ ones is alike. The $\Omega$ and $\kappa$ as a function of $R$ for each plot are numerically determined by the annularly-averaged angular frequencies of all particles at $R$, also time-averaged in the indicated time window.

\begin{figure*}
    \centering
    \includegraphics[width=18.0cm]{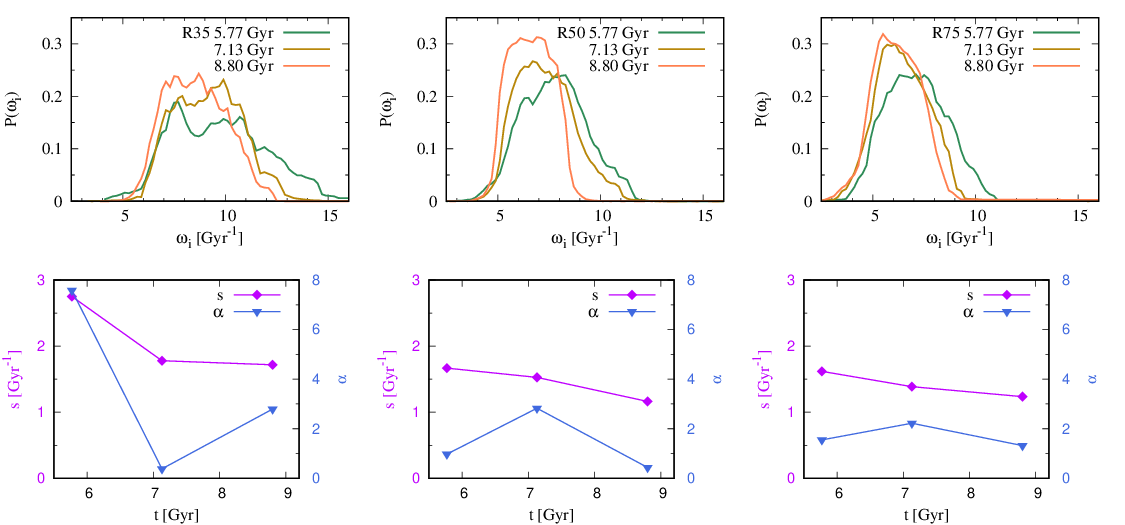}
    \caption{\textit{Top row}: Probability distribution functions of the angular frequencies of the trapped particles $P(\omega_{i})$ for R35 (left panel), R50 (middle panel), and R75 (right panel) at different times (see text for the description of the trapped particles). \textit{Bottom row}: standard deviation $s$ (plotted with the left axis in magenta color) and skewness $\alpha$ (plotted with the right axis in blue color) for the cases on the upper panels at the indicated times.}
    \label{fig:histo_omega}
\end{figure*}

For R35, we observe the buildup of the swing-amplified three-armed modes in the first snapshot, around $1-2$ Gyr, before they are fully established in the following snapshot around $5$ Gyr, at which $A_{3}$ still remains around $0.2$. These swing-amplified modes are characterized by the wavelets reflected between the inner and the outer Lindblad resonances (ILR and OLR) around the corotation radius, as seen by the three components along the $\Omega$ line that are confined within $\Omega\pm\kappa/3$. The limitation by the Lindblad resonances can be explained that the spiral modes are scattered at those spots, weakening the spiral modes and heating the disk as a consequence. That the swing amplification supports the $m>2$ modes is also justified by Fig. \ref{fig:a2} in which $A_{3}$ and $A_{4}$ are the two fastest-growing in the first $3 \ \text{Gyr}$. In the last snapshot which is around the end, the trace of the three-armed modes are still visible although $A_{3}$ is much weakened. On the other hand, the swing-amplified $m=2$ modes are absent. Rather, we observe the buildup of the rigid bar modes of frequency $\sim 8 \ \text{Gyr}^{-1}$ from the center that merely reaches $5 \ \text{kpc}$ in the second snapshot and becomes slightly longer than $5 \ \text{kpc}$ at the end. 
With a closer inspection, the rigid modes are not supported by the corotation resonance until the end as the bar does not reach the $\Omega$ line. It implies that the linear bar modes, albeit their presence since the initial state, are limited by the strong shearing. 
Thus, the principal mechanism underlying the bar growth is not the resonance nor the swing amplification, whose dominant arm number is above $2$.
To resolve this, the evolution of the probability distribution functions of the angular frequencies of the trapped particles $P(\omega_{i})$ are plotted in the top row of Fig. \ref{fig:histo_omega} for the three cases in the last $3 \ \text{Gyr}$.
Particles are considered trapped if their angular frequencies $\omega_{i}$ lie within $\omega_{R}\pm 0.25\omega_{R}$, where $\omega_{R}$ is the angular frequency at $R$ calculated from $\phi_{2}(R)$, and their distances from the axis of $\omega_{R}$ are less than $3\ \text{kpc}$. The particle trapping by the bar potential is affirmed for R35 as the distribution becomes steeper with time, indicating a more organized motion with the bar pattern speed as $A_{2}$ grows. 
While the original analysis of \citet{lynden-bell_1979trap} did not explicitly estimate the characteristic dynamical timescale of the particle trapping, we can infer it from the composite frequency (\ref{eq:omega_i}) used by that work to analyze the orbital mechanics. The period associated with that frequency around the bar end is equal to $3 \ \text{Gyr}$, which is longer than the characteristic timescale of the swing amplification that is $\sim 2 \ \text{Gyr}$ as inferred from the numerical results in Sec. \ref{ssec:evo_morpho}. The latter timescale can be deduced from the works of \citet{julian+toomre_1966,michikoshi+kokubo_2016} who demonstrated that the swing-amplified perturbations evolved in the pulse form that reached the peak after multiple epicyclic periods. More specifically, if the epicyclic period at $10 \ \text{kpc}$, i.e., approximately the extent of the spiral structure, is equal to $0.45 \ \text{Gyr}$, the peaked multi-arm strength at $\sim 2 \ \text{Gyr}$ implies that the timescale of the development of the swing-amplified spiral arms is $4-5$ epicyclic periods. The estimated longer particle trapping timescale is in coherence with the numerical result as the late continual growth of $A_{2}$ lasts longer than $3 \ \text{Gyr}$.
Because the bar development is attributed primarily to the particle trapping by the bar potential, we name such scenario the \textit{dynamics-based bar formation}.

The formation scenario for R75 totally differs from the process in the R35 case because both the linearly unstable bar ($m=2$) and multi-arm ($m=3$) modes of uniform frequencies emerge in the early stage, in agreement with the $A_{m}$ plot in Fig. \ref{fig:a2} and the surface density in Fig. \ref{fig:snap}, before $A_{2}$ exerts the dominance at the end.
Although reducing the halo concentration suggests the shift-down of the preferred swing-amplified arm number, rigid modes spotted in the $m=2$ spectrogram are clearly not from the swing amplification. Firstly, they are established well before $1 \ \text{Gyr}$, while the swing-amplified modes for R35 become prominent well after $2 \ \text{Gyr}$. Secondly, $A_{2}$ reaches a higher peak around $0.7$ before it relaxes, compared with the dominant $A_{3}$ and $A_{4}$ by the swing amplification for R35 that are $\sim 0.4$ at most. A faster growth and a higher reach of the linear $m=2$ modes can be understood by their exponentially growing nature which implies that it is the run-away process. 
The timescale of the growth of the linear modes can be inferred directly from the perturbative form (\ref{eq:d_sigma}) which suggests that the period of the bar pattern speed is the characteristic timescale of the exponential growth. The corresponding period, as drawn from the frequency of the bar modes in the spectrograms, is $\sim 1 \ \text{Gyr}$. It is shorter than the swing amplification timescale, which is around $2 \ \text{Gyr}$, and the particle trapping timescale that is longer than $3 \ \text{Gyr}$.
It is therefore reasonable to deduce that the linearly unstable modes predominantly govern the growth of all non-axisymmetric modes during the first $1 \ \text{Gyr}$, possibly due to a mild shearing. 
These fast-growing modes trigger the corotation resonances since the early time and these resonances are retained until the end despite a violently evolving $A_{2}$. Considering the evolution of $P(\omega_{i})$, it is more steady in the last two snapshots, in accordance with the saturated $A_{2}$ in Fig. \ref{fig:a2}. It indicates that the particle trapping is not as important as in the R35 counterpart because the strong fast-growing bar modes already trap much particles forming the bar since the early time. 
As the resonated linearly unstable bar modes principally take the responsibility to constitute a bar, with an absence of the swing amplification and a negligible role of the particle trapping, this bar formation scenario can be described as the \textit{linear-mode-based bar formation}.

The bar formation and evolution characteristics for R50 share similarities with parts of the two regimes above. We observe the rapid rise of the linear three-armed modes of frequency $\sim 9 \ \text{Gyr}^{-1}$ which become swing-amplified in the next snapshot before they are weakened at the end. The switch from the linear to the swing-amplified features can be interpreted that none of the two is completely dominant, so they emerge according to their characteristic timescales. On the contrary, the $m=2$ linear modes are absent in the first snapshot. We rather observe the buildup of the swing-amplified $m=2$ modes between $\Omega\pm\kappa/2$ before the bar of frequency $\sim 7 \ \text{Gyr}^{-1}$ in corotation resonance is fully developed in the last snapshot, which looks like that established in R75.
The evolution of the $m=2$ spectrogram is more complex than the two cases before. It indicates that the swing amplification precedes and triggers the linear modes. This can be explained by the moderate concentration degree which leads to two important outcomes. The reduced halo concentration from that of R35 shifts the dominant swing-amplified arm number toward $2$, as seen by Fig. \ref{fig:a2} in which $A_{2}$ is one of the two fastest-growing modes in place of $A_{4}$. On the other hand, a moderate shearing degree enables the development of rigid linear modes for some $m$ at the early time. This explains the early emergence of the linear $m=3$ modes before the swing amplification comes into play. The emergence of the $m=3$ linear modes hints the possibility of the involvement of the $m=2$ counterparts in developing the bar, seeded by the swing amplification.
The $P(\omega_{i})$ plot and the late growth of $A_{2}$ indicate that the particle trapping also takes a part in developing a bar, but it is not as important as in the R35 counterpart. In other words, the bar formation for R50 is not predominantly regulated by a single process. All the focused mechanisms are involved in forming a bar. The combined constructive processes are justified by a higher $A_{2}$ at the end compared to those of R35 and R75.

In order to better quantify the evolution of the shape of $P(\omega_{i})$ to delineate the dynamics-based bar formation scheme from the other, the probability distribution is fitted with the skew normal distribution
\begin{equation}
f(\omega_{i})=\frac{1}{\sigma \sqrt{2\pi}}e^{-\frac{(\omega_{i}-\omega_{0})^{2}}{2\sigma^{2}}}(1+\text{erf}(\alpha(\frac{\omega_{i}-\omega_{0}}{\sigma\sqrt{2}}))),
    \label{eq:skew}
\end{equation}
where $\text{erf}$ is the error function, $\alpha$ is the skewness, $\omega_{0}$ is a constant relating with the mean, and $\sigma$ relates with the standard deviation $s$ as
\begin{equation}
s^{2}=\sigma^{2}(1-\frac{2\alpha^{2}}{\pi(1+\alpha^{2})}).
    \label{eq:variance_skew}
\end{equation}
As a result, $s$ and $\alpha$ at the different times for the three cases are plotted in the bottom panels of Fig. \ref{fig:histo_omega}. For the R35 plot, the transition from a highly asymmetric and dispersed distribution of $\omega_{i}$ at $5.77 \ \text{Gyr}$, designated by $s \sim 3 \ \text{Gyr}^{-1}$ and $\alpha\sim 8$, to a steeper and more symmetric $P(\omega_{i})$ in the following times, representing a more ordered motion of particles constituting a bar, is spotted. The stage of a more coherent motion is described by $s$ and $\alpha$ that are well below $2 \ \text{Gyr}^{-1}$ and $3$, respectively. On the other hand, both parameters for R50 and R75 exhibit a mild decrease with time on average and they are always below those values. It justifies a less significant role of the particle trapping in forming a bar.

\subsection{Dynamical perspectives of bar formation sub-mechanisms and relevance to past works}
\label{ssec:discuss_process}

A full inspection of the Fourier spectrograms of multiple modes and the particle angular frequency profiles helps us to better understand the different bar formation scenarios involving different sub-mechanisms. The swing amplification was often related with the bar formation while our results further clarify that the its involvement is complex than initially understood. First of all, the exponentially growing linearly unstable modes and the swing amplification are the two distinct entities and can be involved in different roles in forming a bar depending on the preferred mode numbers of each mechanism. Because the linearly unstable modes always prefer $m=2$, the preferred number of arms from the swing amplification, that varies with the halo concentration, is the key factor in determining the formation scheme. If the dominant swing-amplified arm numbers are well above $2$, i.e., the R35 case, the two mechanisms react counter-actively. Owing to a strong shearing, the multi-arm swing-amplified modes limit the growth of the linear bar modes to the point that the latter modes cannot reach the corotation radius, albeit the decay of the spiral arms and the continual growth by the mechanical particle trapping. This is possibly the scenario found in \citet{sellwood+carlberg_1984} as the $m>2$ modes towered above the bar modes at the beginning but the bar modes dominated afterward. More refined analyses of similar disks in continuation by \citet{sellwood+carlberg_2014} described the short bar as the swing-amplified modes reflected off the center. From our results, we may have a different view on the bar formation mechanism. These growing modes do not resonate at all and develop relatively slowly compared with other cases that involve the resonances; the $A_{2}$ plot suggests that the bar growth lasts more than $5 \ \text{Gyr}$, which is longer than the time for the establishment of the resonated swing-amplified modes, but the resonance is still not established. We conclude that it is regulated by the particle trapping. 
It is possible that this scenario happens only when $Q$ is close to $1$. This is because if $Q$ is much greater than $1$, the linear bar modes are greatly suppressed, rendering more long-lasting swing-amplified multi-arm modes and making the formation of the bar in such high-$Q$ condition unlikely. Disks with $Q\geq 1.2$ were rather subject of studies of the long-lasting multi-arm modes \citep{baba_et_al_2013,sellwood+carlberg_2019,sellwood+carlberg_2021}.

For R50, the dominant swing-amplified arm number approaches $2$. The Fourier spectrograms suggest that the swing amplification works constructively with the linear bar modes. With the particle trapping in addition, the bar of R50 is stronger than that in R35. The linear bar modes seeded by the two-armed swing amplification could possibly be the situation in \citet{donner+thomasson_1994,sellwood_2011,roca-fabrega_et_al_2013,baba_2015,saha+elmegreen_2016} as the bi-symmetric structures predominantly appeared along the evolution and the bars were gradually developed without a signature of early fast-growing bar modes. We note that this scenario could occur for a higher $Q$ as in those works, their $Q$ could exceed $1.3$. 
The R75 case forms a bar solely owing to the linear modes stemming from the noises, independently from the swing amplification. A bar as a result of the linearly unstable bar modes could be the case for \citet{minchev_et_al_2012,sellwood_2016cmc,saha+elmegreen_2018},
as evaluated by the strongly-growing pattern of the bar amplitude since the start and the spectrogram. A notable remark by \citet{sellwood_2016cmc} speculated the different bar-forming mechanisms from the linear bar modes or the swing amplification in halos of different concentrations, as we also speculate. That speculation is partly true as we find a more complex interplay between those two mechanisms that can be counter-active, i.e., the R35 case, constructive which is the R50 case, or independent as for R75. Especially for the R35 case, neither the swing amplification nor the resonating linear bar modes is responsible for the bar growth. It is the particle trapping that underlies the bar formation, in continuation of the restrained bar modes after the multi-arm modes decay.

The sensitivity of the engendered arm number to the prescribed modes of perturbations was also an active topic of investigation. Some studies introduced the $m$-mode perturbations during the disk evolution by means of, for examples, restricting the mode number of the Fourier-expanded force calculations \citep{sellwood+kahn_1991,sellwood_2012,de_rijcke_et_al_2019} or introducing $m$ point masses  \citep{kumamoto+noguchi_2016,sellwood+carlberg_2021}. They found that the awakened mode number correlated with the mode number of the perturbers. In our study, the perturbations embedded in the initial states are purely Poissonian and we employ the tree code that does not segregate the force into modes, thus the resulting preferred mode numbers and the different processes that occur are the direct consequences of the true interplays between mechanisms, which reflect more appropriately the dynamics of the real disk galaxies.

That the two-armed modes are not always generated by the swing amplification may render more complexities for the estimate of the dark matter halo profile. A well-known method proposed by \citet{athanassoula_et_al_1987} to derive the halo configuration parameters was based on the hypothesis that the two-armed spiral modes were swing-amplified. We demonstrate that the two-armed modes can also arise in a further diluted condition but they originate only from the linearly unstable modes, which we specify to be a distinct entity from the swing amplification. Therefore, a more correct estimate of the halo parameters should take the fact that there are a number of mechanisms generating the bar and the grand-design spiral pattern into account. The post-formation morphologies can be indicative to the bar evolution track as we demonstrated in Sec. \ref{ssec:evo_morpho}.

The different mechanisms underlying the bar formation for different halo concentrations can also explain the different bar environments in Fig. \ref{fig:snap} and configurations of $\tilde{A}_{2}(R)$ in Fig. \ref{fig:a2b2r}. For R75, the non-vanishing bi-symmetric modes beyond the corotation radius are the remnants of the fast-growing bar modes earlier. A more gentle shearing allows these modes beyond the bar, limited by the corotation radius \citep{contopoulos_1980}, to remain. These explain the highly eccentric isodensity contours of the mass surrounding the bar and the significant level of $\tilde{A}_{2}(r)$ beyond the bar. On the contrary, the fast and extensive bar growths are limited in the disks with high and moderate shearing such as R35 and R50, so the bar modes cannot reach a great radial distance. It is justified by $\tilde{A}_{2}(R)$ that is near zero beyond the bar end. The bar for R35 is particularly short and slowly developed without the resonance, thus the bi-symmetric modes do not exist at all beyond the bar, leaving the mass surrounding the bar close to the circular symmetry.

Considering the role of the ILR, the fact that the ILR is dissipative by nature suggests that it frames the presence of the linear modes from the inside. In other words, bar modes are permitted to grow only in the cavity between the ILR and the corotation resonance \citep{contopoulos+papayannopoulos_1980,sparke+sellwood_1987}. This basis allows a stabilization process against bar modes by including a massive bulge in the disk center. A massive bulge introduces a prominent peak of the $\Omega-\kappa/2$ line in the inner region which leads to two important effects. It shifts the ILR outward from the center and narrows the cavity, both of which hinder the development of the elongated global bar modes so that the disk develops the swing-amplified spiral arms in place \citep{donner+thomasson_1994,saha+elmegreen_2018}. With regard to our bulge-less model, varying the concentration of the halo does not significantly narrow the cavity. We demonstrate that a wide cavity does not always promote the bar development in the same manner. Depending on the halo concentration, the bar modes stem from different mechanisms.
%
%
Lastly, it is true that we adopt the bulge-less model, whereas a real galaxy may consist of a bulge in addition. The halo-to-disk mass and size ratios of the R35 case suggests that the central mass concentration can be comparable to that of the moderately-concentrated halo with a bulge (see, for instance, \citealt{saha+naab_2013,athanassoula_et_al_2013tri,jang+kim_2023,bland_hawthorn_et_al_2023}). Thus, the dynamics-based bar formation can be the case for other galaxies if their central mass concentration and $Q_{min}$ fall in the range of those of the R35 case.

\subsection{Interactions between the non-axisymmetric modes and the surrounding}
\label{ssec:heat_dlz}

\begin{figure*}
    \centering
    \includegraphics[width=18.0cm]{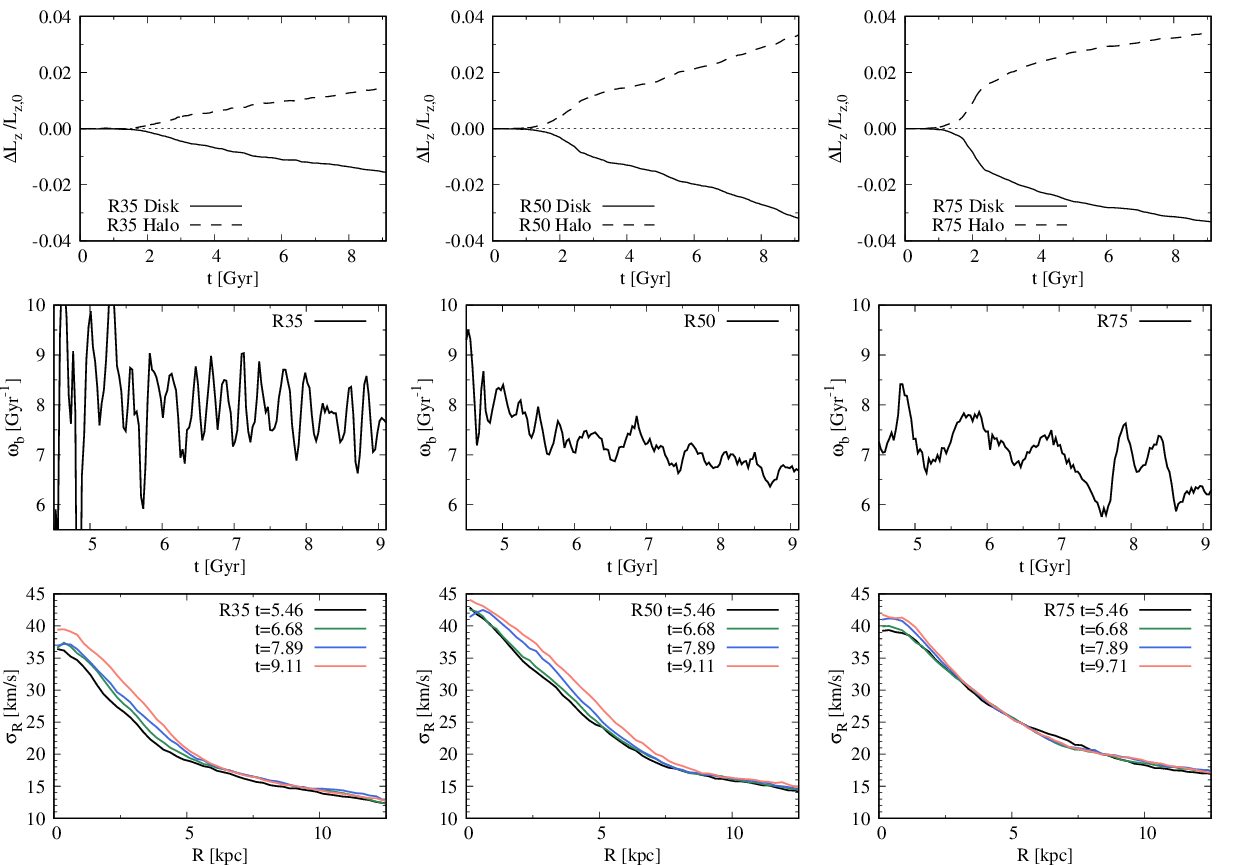}
    \caption{\textit{Top row}: Changes of the disk (solid lines) and the halo (dashed lines) $z$-axis angular momenta relative to the initial total angular momentum $\Delta L_{z}/L_{z,0}$ as a function of time for different cases. \textit{Middle row}: Bar pattern speed $\omega_{b}$ as a function of time, calculated from the rate of change of the bar phase $\phi_{2,tot}$, for different cases. \textit{Bottom row}: Radial velocity dispersion as a function of radius $\sigma_{R}^{2}$ for different cases at the indicated times.}
    \label{fig:moment_pat_dispr}
\end{figure*}

In this section, we examine the angular motion of the bar modes and the interactions with the disk environment and the halo. To this aim, the changes of the $z$-axis angular momenta of the disk and the halo in units of the initial total one $\Delta L_{z}/L_{z,0}$; the bar pattern speed $\omega_{b}$, calculated from the rate of change of the total bar phase $\phi_{2,tot}$; and the radial velocity dispersion profile $\sigma_{R}(R)$ for R35, R50, and R75 are shown in Fig. \ref{fig:moment_pat_dispr}. The disk-to-halo angular momentum transfer is affirmed by the decay of the disk angular momentum with time in all cases, in coherence with the angular momentum gain of the halo. With a closer inspection, the exchange amount for R35 is significantly lower than those for the two cases, although the bar is continually growing. This is because the bar evolution is not supported by the resonance, which plays a central role for the angular momentum transfer \citep{athanassoula_2003}. The absence of the resonance also manifests in the modestly decreasing $\omega_{b}$ for R35, whereas those for R50 and R75 substantially decay owing to the a more pronounced angular momentum transfer. This finding can be important as the bar evolution has regularly been associated with the slowdown, but we hereby demonstrate that a bar does not necessarily slow down remarkably, even less than $0.5 \ \text{Gry}^{-1}$ in the course of $4 \ \text{Gyr}$ if it is established without support from the resonance, i.e., the dynamics-based mechanism. The quasi-steadiness of the bar pattern speed has indeed been documented but those studies concluded in the different factors such as the halo fast rotation \citep{long_et_al_2014,collier+madigan_2021,li_et_al_2023,chiba+kataria_2024,jang+kim_2024} or the high gas fraction \citep{villa-vargas_et_al_2010,beane_et_al_2023}. In the mentioned works, the Fourier spectrograms were not inspected, thus it could not be definitely concluded whether the slow decrease of the bar pattern speed was attributed to the absence of the resonance. The correlation between the absence of the resonance and the low angular momentum transfer was concluded elsewhere by \citet{villa-vargas_et_al_2009,collier_et_al_2019}, via the spatio-temporal analysis of the angular momentum loss, but their consideration was limited to the $m=2$ modes. Our study provides more complete details from the Fourier spectrograms of many modes which lead to the results that, firstly, the resonance is really absent for the many modes in most of the time. Secondly, albeit the absence of the resonance, there is still the long-lived short bar spotted in the spectrogram. This bar formation and evolution scenario can be the explanation of the existence of the short fast bar (or the nuclear bar) spotted in observations \citep{wozniak_2015,sierra_et_al_2015,buttitta_et_al_2023,ghosh_et_al_2024}. Not only the strong shearing, the similar fast bar without the resonance support was also found in another environment unfavorable to the formation of the extensive rigid bar modes, which is the counter-rotating halo \citep{lieb_et_al_2022}.

The $\sigma_{R}(R)$ plots, however, demonstrate the opposite tendency as the radial heating degree, evaluated by the increase of $\sigma_{R}$ within $5 \ \text{kpc}$, for R35 is more pronounced from $5.46-9.11 \ \text{Gyr}$ than the two other cases. Unlike the angular momentum transfer, the resonance between the non-axisymmetric modes and the disk natural frequencies is not a requisite for the radial heating. The agent of the heating is simply the potential of the non-axisymmetric components of the disk such as the bar, the spiral arms, the clumps, or even the combination of them.

The physical and kinematical properties of the bars formed in different conditions were active subjects of investigation by many research groups. A series of works in 1990s \citep{masset+tagger_1997,rautiainen+salo_1999,rautiainen+salo_2000} examined the bar kinematics in various disk systems, exemplified via the $m=2$ spectrograms. Their principal focus was on the coupling between modes such as the bar, the ring, and the spiral arms, with the disk, while the explanation of the origin of the bars originating from different systems was not much considered. Although the swing amplification and the bar modes were mentioned in \citet{masset+tagger_1997}, the distinction between the two things was not addressed. Our study clarifies that the two processes are the distinct entities that can act differently in forming a bar depending on the disk environment.
The work of \citet{roca-fabrega_et_al_2013} considered cases with various central mass concentrations and those cases could be classified into the disks that rapidly formed the barred-spiral structure and the disks that were bar-stable, yielding the spiral arms in place. We address the points that we find unclear as follows. The Fourier spectrogram at the time the bar is fully established is not sufficient in pursuing the evolutionary track. This is demonstrated by R50 and R75 that exhibit different bar-forming procedures but they lead to a similar bar in the spectrogram. 
The other work worth attention was \citet{pettitt+wadsley_2018} that generated the initial disk systems with different shapes of the rotation curves, acquired by differentiating halo parameters. As a consequence, bars with various strengths, lengths, and pattern speeds, were reported. Although their spectrograms of various bars might be indicative to either the short fast-rotating bar and the resonance-supported bars, their dynamical origins were not much investigated. In this study, we are able to spell out at least $3$ different scenarios for the bar formation.

\section{Analysis of interplay between growing bar modes and swing-amplified spiral modes}
\label{sec:shear_stab}

\begin{figure}
    \centering
    \includegraphics[width=8.0cm]{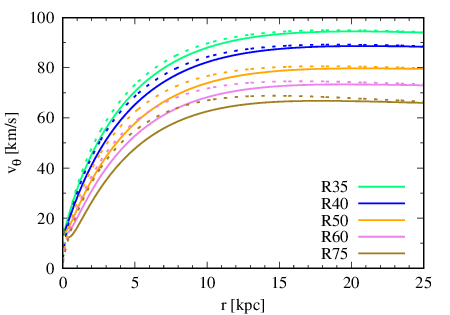}
    \caption{Rotation curves derived from the Jeans equation (\ref{vthetamean}) for all cases, plotted in the solid lines, in comparison with the circular velocities of the sames cases calculated directly from the composite potential, plotted in the dashed lines with the same colors.}
    \label{fig:rot_curve}
\end{figure}

The transition of the scenario of the bar formation between the two extremities of $r_{h}$, reported at length in Sec. \ref{sec:nume}, can be better understood by the following analysis of the kinematical structure of the different disks. We speculate that the factor determining which path the bar-forming disk follows is the counteraction between the shearing, a driving factor of the swing amplification, and the rigid modes, as a representative of the fast-growing linear bar modes. For this purpose, we define the shearing-to-rigid ratio $\Xi$ as a function of radius to be
\begin{equation}
\Xi =  \frac{2\varpi(\bar{v}_{\theta}-\varpi R)}{\varpi^{2}R},
    \label{shear-to-rigid_eq}
\end{equation}
where $\bar{v}_{\theta}$ is the rotation curve obtained from the Jeans equation (\ref{vthetamean}). The variable $\varpi$ is the pattern speed of the rigid mode and we consider it as the constant global rigid mode frequency. The numerator in Eq. (\ref{shear-to-rigid_eq}) stands for the Coriolis force seen at $R$ from the rotating frame of angular frequency $\varpi$, which represents how forcefully the disk mass at $R$ is pulled out of the rigid frame. The denominator, opposingly, is the norm of the centrifugal force, denoting the strength of the force framing the rigid motion. This ratio represents the comparative degree of the shearing to the rigid motions in the disk, on which we speculate to be the important factors. By definition, the direction of the Coriolis force is either along or opposite to the centrifugal force from $\varpi$. If it is present, it deviates the mass fraction at $R$ from the coherent motion of frequency $\varpi$.

\begin{figure*}
    \centering
    \includegraphics[width=18.0cm]{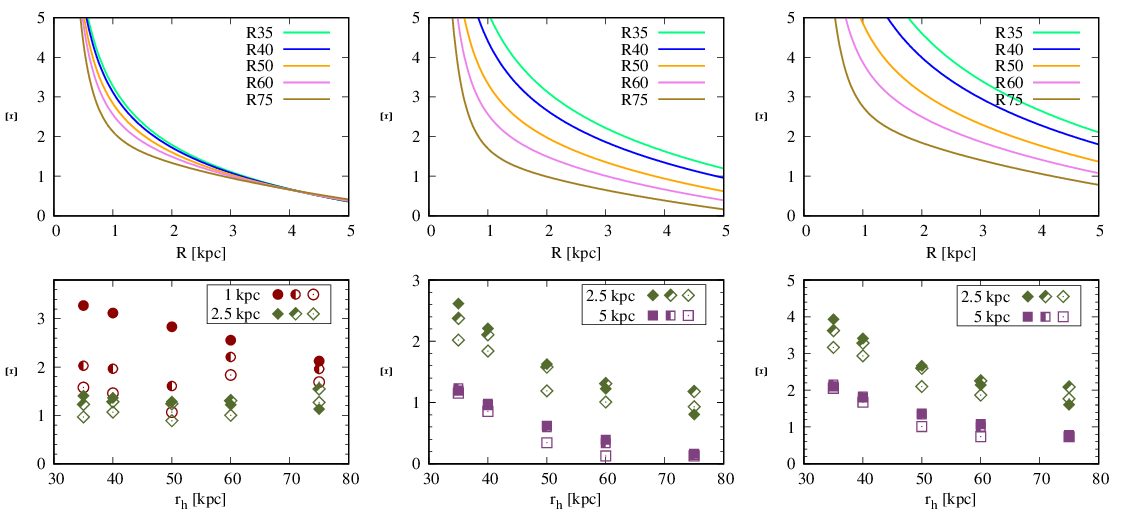}
    \caption{\textit{Top panels}: The shearing-to-rigid ratio $\Xi$ as a function of radius $R$, computed from the initial rotation curve, for different disks with the bar pattern speeds $\varpi$ equal to the mass-weighted average angular frequencies from the entire disk (left panel), $9 \ \text{kpc}^{-1}$ (middle panel), and $7 \ \text{kpc}^{-1}$ (right panel). \textit{Bottom panels}: The same ratio plotted as a function of $r_{h}$ taken at different indicated radii, calculated from $\varpi$ as for the panels above. The plots exhibit $\Xi$ calculated from the rotation curves at the initial time (filled points), during $4.10-5.01 \ \text{Gyr}$ (half-filled points), and during $8.20-9.11 \ \text{Gyr}$ (empty points).}
    \label{fig:ratio_shear_rigid}
\end{figure*}

The rotation curves of all cases derived from the Jeans equation (\ref{vthetamean}) in comparison with the circular velocities computed directly from the composite potential of the $N$-body model are plotted in Fig. \ref{fig:rot_curve}. The effect from $Q$ becomes more minor as the concentration increases, which can be understood that the dynamical equilibrium is more rotationally-supported and requires a less degree of support from the random velocity via $Q$.
We first of all consider $\Xi$ as a function of $R$ for different $\varpi$ and $r_{h}$ calculated from the initial rotation curves, that are shown in the top panels of Fig. \ref{fig:ratio_shear_rigid}. We choose $\varpi$ to be the self-consistent mass-weighted average of the angular frequency over the entire disk (left panel), $9 \ \text{Gyr}^{-1}$ (middle panel), and $7 \ \text{Gyr}^{-1}$ (right panel), as inferred from the typical magnitudes of the bar pattern speeds in Sec. \ref{ssec:heat_dlz}. 
Although $\varpi$ is chosen differently, the figures exhibit the generic $\Xi$ curves of the order of unity up to a disk scale radius that decrease with $R$ and shift downward accordingly with increasing $r_{h}$, as expected from a lessened shearing strength. The comparative degree of the shearing and the rigid modes is better exemplified when we adopt a fixed $\varpi$ while a self-consistently averaged $\varpi$ yields the curves for different $r_{h}$ that are less distinguishable. It is because the self-consistently averaged $\varpi$ varies significantly with $r_{h}$, ranging from $8-12 \ \text{Gyr}^{-1}$ depending on $r_{h}$, while the true pattern speeds for different cases, as inferred from Fig. \ref{fig:moment_pat_dispr}, differ subtly as they are in the range of $7-8 \ \text{Gyr}^{-1}$. Furthermore, the average $\varpi$ significantly overestimates the true pattern speed that results in less logical curves.

We further investigate $\Xi$ calculated at some selected $R$ and the results are depicted as a function of $r_{h}$ in the bottom panels of Fig. \ref{fig:ratio_shear_rigid}. The ratio is computed in the three time windows: $0 \ \text{Gyr}$, $4.10-5.01 \ \text{Gyr}$, and $8.20-9.11 \ \text{Gyr}$, which represent the initial, the middle, and the end times of the dynamical evolutions. The plots for the later times are for the investigation of the robustness of the ratio $\Xi$ while the disk evolves.
It turns out that using the self-consistently averaged $\varpi$ in determining $\Xi$ cannot reflect the actual dynamical process of the bar formation properly, in coherence with the plots above. The variation of $\Xi$ with $r_{h}$ is evident only in the inner region, namely, at $1 \ \text{kpc}$, at the initial time. Afterward, the decrease with $r_{h}$ is no longer observed. It is explainable because the dynamics of this part is subject to numerical artifacts such as the low particle number, as well as physical factors such as the dissipative nature of the ILR close to the center. At $2.5 \ \text{kpc}$, although $\Xi$ is reasonably robust over time, the ratio is saturated while we vary $r_{h}$. 
On the other hand, using a fixed $\varpi$ yields a better agreement with the numerical results. The ratio $\Xi$ decreases with $r_{h}$ in coherence with the transition of the bar formation scheme from the dynamics-based bar formation in a strong shearing background to the linear-mode-based bar formation in a weakly shearing background. The ratio is reasonably robust over time until the end if we measure it well outside the disk center, at $2.5$ and $5 \ \text{kpc}$ as we perform, although the disk evolution manifests structural and kinematical changes with time involving processes such as the disk heating, the angular momentum transfer, and the bar slowdown (see Sec. \ref{ssec:heat_dlz}). This emphasizes the validity of all theoretical frameworks that regard the bar and the non-axisymmetric modes as the perturbative components in a dynamically equilibrated disk.

The formulation of this simple parametric ratio to describe the different bar formation processes provides a new insight into the topic of the bar instability as in past studies, the consequence of the shearing has always been directed along the other lines.
In the morphological aspect, the initial shearing degree was frequently associated with the pitch angle of the spiral arms \citep{grand_et_al_2013,michikoshi+kokubo_2016,yoshida+kokubo_2021}. 
For a more relevant scope, the differential rotation was related with the Coriolis force by \citet{baba_et_al_2013}. It was considered as a factor damping the growth of the spiral structure when it exceeded the local gravitational binding force from the spiral perturbation. We continue with those ideas but we reconsider the Coriolis force from the shearing as an opposing factor to the rigid modes in determining the bar formation scheme.

Despite the improved view on the bar formation mechanism by our analysis on $\Xi$, there are still rooms for improvement. Firstly, the magnitude of $\Xi$ is sensitive to the choice of $\varpi$, despite that the $\Xi$ curves vary generically. Thus, to pinpoint the precise critical $\Xi$ delineating the two regimes is not practical now. Secondly, we consider the rotation curves derived from the Jeans equation in the low-$Q_{min}$ regime, which is not far from the nearly circular-orbit limit. If $Q_{min}$ is elevated, high velocity dispersion gives rise to the non-negligible pressure force which has the other crucial roles such as to prolong the bar formation or to stabilize the disk. In other words, for such analysis to cope with the elevated range of $Q_{min}$, the effect from the pressure must be considered.

\section{Conclusion}
\label{sec:conclu}

We investigate the bar formation process in the disk-halo systems with various halo scale radii $r_{h}$, in the attempt to varying the central mass concentrations. We scope on the $Q_{min}$ close to $1$ to ensure that the bar instability manifests within the simulated timescale for all $r_{h}$ so that the underlying formation process and the post-formation evolution can be scrutinized. 
The morphological and kinematical analyses are based on the three milestone theories: the linear theory, the swing amplification, and the orbital or particle trapping. The main objective is to better understand how these three mechanisms are involved in the bar formation in the different cases.
In the high-concentration regime, the bar formation is the secondary process after the early dominance of the multi-arm modes as a result of the swing amplification. A strong shearing promotes the preferred numbers of arms higher than $2$ and limits the growth of the linear bar modes. After the multi-arm modes decay, the bar can be developed to the full bar by the particle trapping mechanism. From the beginning to the end, the corotation resonance is not involved at all, so the bar is formed mechanically, i.e., the dynamics-based bar formation. The absence of the resonance is in line with a low amount of the disk-halo angular momentum transfer and a mild decrease of the bar pattern speed. 
On the contrary, a disk in a halo of low concentration exhibits the linear-mode-based bar formation because the bar is the product of the fastest-growing bi-symmetric linearly unstable modes of uniform frequency. That fast growth prompts the corotation resonance since the early time and such resonance backs the saturated bar until the end, in accordance with the pronounced disk-halo angular momentum transfer and bar slowdown. The particle trapping in enhancing the bar strength and the swing amplification are not as important as in the regime before. The disk in the in-between halo concentration exhibits mixed, but identifiable, features from both regimes in the way that the swing-amplified two-armed modes, which become the preferred modes by reducing the halo concentration, arise first before that bi-symmetric swing-amplified product triggers the resonance-supported linear bar modes that underlies the bar evolution until the end.

In connection with the observations, the bars formed in halos of different concentrations can be distinguished by the full morphological and kinematical analyses. In the first aspect, a bar built in a high-concentration halo is relatively short and it is surrounded by a dense circular background with vanishing bi-symmetric Fourier amplitude. For the low-concentration regime, on the contrary, the bar modes are the fastest-growing entities in the disk center, towering above all multi-arm modes since the beginning. Thus, there are remnants of the two-armed modes beyond the bar, specified by the non-vanishing bi-symmetric Fourier amplitude and the elliptical bar environment. To differentiate one from the other in a more specific way, the ellipticity ratio of the different isodensity lines and the radial profile of the Fourier bar amplitude can be of use.
The differentiation between the bars with and without resonance in the kinematical map can be performed by the Fourier spectrogram, which is deducible in the case of the numerical simulation. In observations, this can be carried out by methods of morpho-kinematical analyses combined. Not only the specification of the bar formation processes taking place at the two extremities of the concentrations, full analyses of those profiles also prove effective in pinpointing the bar formation in the intermediate halo concentration, yielding combined features between those from the two regimes as reported at length in Sec. \ref{sec:nume}.

The transition from the dynamics-based to the linear-mode-based bar formation by increasing $r_{h}$, equivalent to lowering the central mass concentration, can be understood by the reduced shearing strength relative to the strength of the rigid modes. We parameterize such interplay by the ratio of the Coriolis to the centrifugal forces, or $\Xi$. The radial profile of $\Xi$ in the disk interior shifts down with increasing $r_{h}$ in accordance with the numerical results exhibiting the transition of the bar formation scenario. The ratio is proved robust over time if we measure it around the disk scale radius. However, the proposed analysis is only the first step as it solely incorporates the disk self-gravity and we restrict ourselves in the $Q_{min}\sim 1$ limit. A more agitated disk might have the pressure as another important factor and it can manifest more possible evolution tracks such as the slow bar formation or the stable disk against bar modes.

\begin{acknowledgements}

This work is supported by (i) Suranaree University of Technology (SUT), (ii) Thailand Science Research and Innovation (TSRI), and (iii) National Science, Research and Innovation Fund (NSRF), project no. 195242, and the Program Management Unit for Human Resources \& Institutional Development, Research and Innovation (PMUB), grant number B05F640075. Numerical simulations are facilitated by HPC resources of Chalawan cluster of the National Astronomical Research Institute of Thailand.

\end{acknowledgements}


\end{document}